\documentclass{cmspaper}
\usepackage{epsfig}
\begin{document}


\begin{titlepage}

   \date{\today} 

  \title {Type-I Error or Mass Bias ? \\ An Investigation on the $\Omega_b$ Discovery}

  \begin{Authlist}
       T.~Dorigo 
       \Instfoot{INFNpd}{INFN-Padova, Italy}
  \end{Authlist}



  \begin{abstract}
  The D0 and CDF collaborations recently published two independent analyses that both claim to represent the observation of the $\Omega_b$ particle, a baryon made up by a (bss) quark combination. Both signals are estimated to exceed the statistical significance of five standard deviations; however, the mass measurements derived from the candidates differ by over six standard deviations, accounting for estimated systematics. Measured rates also appear to differ, although they remain compatible within the large uncertainties. 

In this paper the author recomputes the significance of the D0 result, showing that it was considerably overestimated in the original publication; he then investigates with a pseudoexperiment-based approach which, among different hypotheses, appears the most likely cause of the observed discrepancy between the D0 and CDF signals.
  \end{abstract} 

  
\end{titlepage}

\setcounter{page}{2}


\section{Introduction}
\vskip .5cm

In the last few years the large statistics of proton-antiproton collisions produced by the Tevatron have allowed the CDF and D0 experiments to shed some light in the largely unknown territory of bottom baryons. The only baryon containing bottom quarks directly observed before the turn of the millennium, the $\Lambda_b$~\cite{pdg}, has been joined by observations of the $\Sigma_b$ and $\Sigma^*_b$~\cite{CDF_sigmab}, the $\Xi_b$~\cite{d0_chib,cdf_chib}, and most recently, the $\Omega_b$, first claimed by D0 in 2008~\cite{d0_omegab}, and then by CDF this year~\cite{cdf_omegab}. The latter two results are in conflict with each other both in mass and production rate. Until larger-statistics measurements will be made available one is left wondering which, among several concurrent hypotheses, is the most likely cause of the discrepancy, and what is the most credible estimate for the $\Omega_b$ baryon mass. This study tries to address the above issues quantitatively using a simple-minded pseudoexperiment approach.

In section 2 the two analyses are briefly summarized. This allows to put in evidence an oversight by D0 in the derivation of the significance of their observation; an independent assessment of the same number is performed with the help of pseudoexperiments, which highlights the impact of the ``look elsewhere effect'', a common problem of searches for signals of unknown mass. Section 3 focuses on the mass determinations of the $\Omega_b$, and on the causes of the observed discrepancy between the CDF and D0 measurements. Some conclusions are offered in section 4.

\section{Observations and significances}
\vskip .5cm

\subsection{The D0 observation}
\vskip .3cm

D0 searched $1.3 fb^{-1}$ of Tevatron Run II $p \bar p$ collisions data for the signal of $\Omega_b$ baryon decays into $J/\psi \Omega^-$ pairs, by matching a $J/\psi \to \mu^+\mu^-$ signal with a fully reconstructed $\Omega^- \to \Lambda K^- \to (p \pi^-)K^-$ decay~\footnote{Particle reactions quoted in this article of course imply charge-conjugate states.}. The analysis used a multivariate selection based on ``boosted decision trees'' to increase the purity of selected events, relying on a Monte Carlo simulation of $\Omega_b$ production to model the signal kinematics, and wrong-sign $J/\psi (\Lambda K^+)$ events found in the data as a model of backgrounds. 

To reconstruct the $\Omega_b$ mass, D0 substituted PDG~\cite{pdg} world-average values to the reconstructed mass of $J/\psi$ meson and $\Omega^-$ baryon candidates, which allowed a significant improvement in the mass resolution of the system: according to the paper, \par

{\em ``We calculate the $\Omega_b$ candidate mass using  the formula $M(\Omega_b) = M(J/\psi \Omega^-) -M(\mu^+ \mu^-) -M(\Lambda K^-) + M(J/\psi) + M(\Omega^-)$. [...] This calculation improves the mass resolution of the MC $\Omega_b$ events from 0.080 GeV to 0.034 GeV.''}

\begin{center}
\begin{figure}
\begin{minipage}{0.47\linewidth}
\epsfig{file=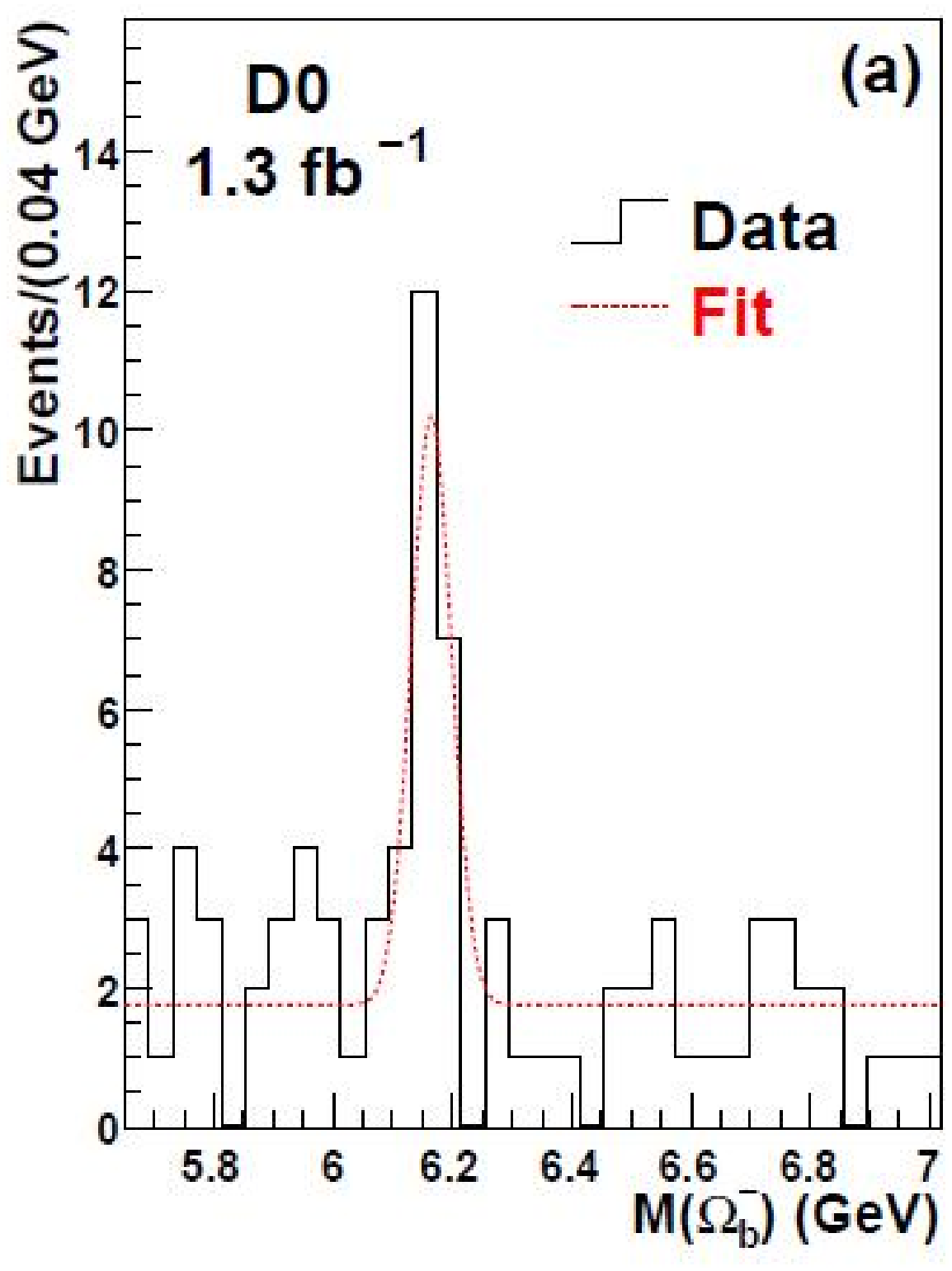, width=6cm, angle=0, clip=}
\end{minipage}
\begin{minipage}{0.47\linewidth}
\epsfig{file=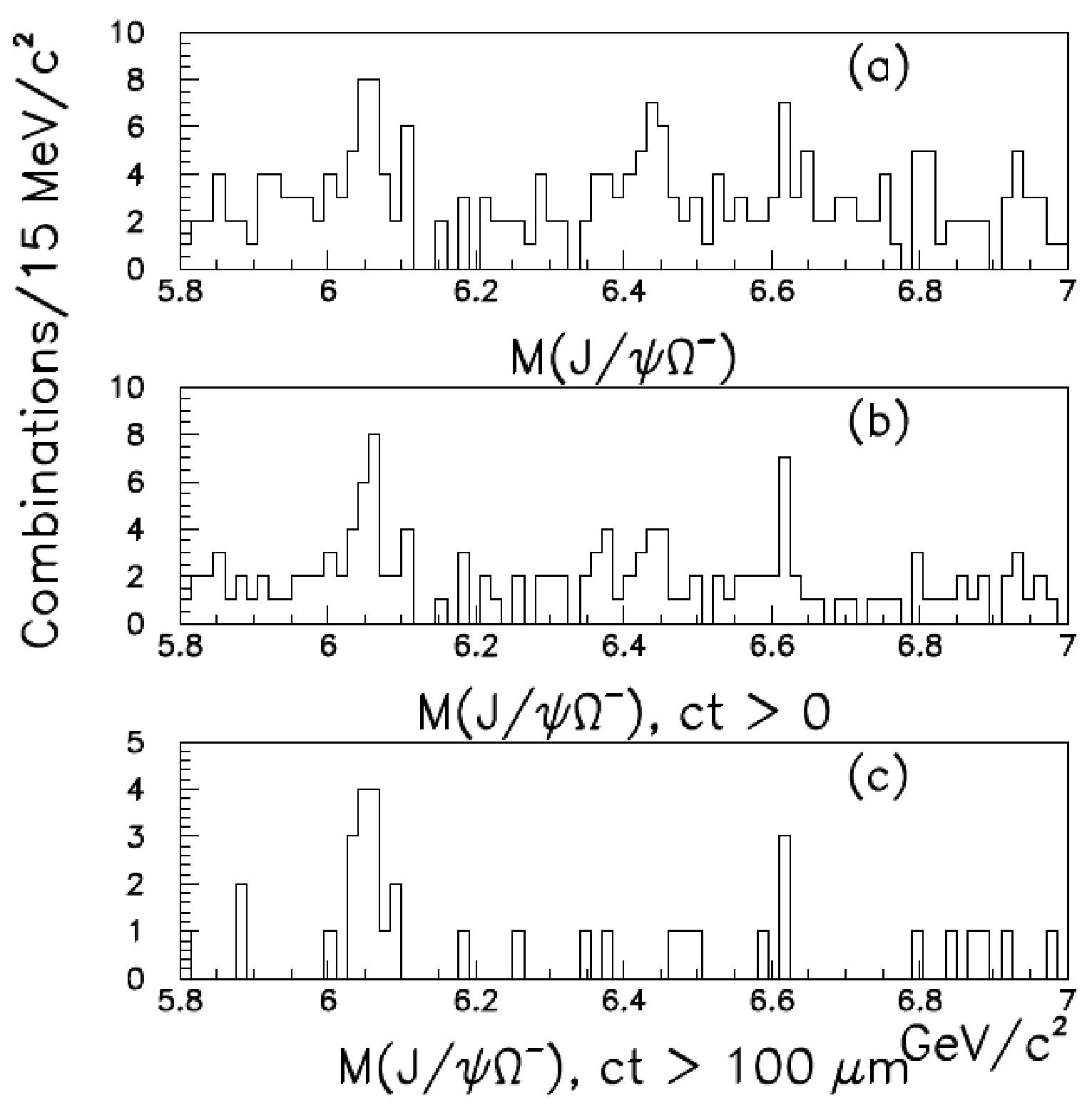, width=7.5cm, angle=0, clip=}
\end{minipage}
\caption{\em Left: The fit to 79 event candidates extracted by D0, which finds a signal of 17.8 events with a claimed significance of 5.4 standard deviations. Right: Mass distribution of $\Omega_b$ candidates extracted by the CDF analysis. Top: all candidates; center: candidates with decay length exceeding $50 \mu m$; bottom: candidates with decay length exceeding $100 \mu m$. The sample with no requirements on the decay length is the default one used in~\cite{cdf_omegab} for a combined mass and lifetime fit. }
\label{f:omegab_signals}
\end{figure}
\end{center}

The selection of event candidates led to a sample of 79 events with reconstructed invariant mass between 5.65 and 7.01 GeV. A unbinned likelihood fit of the distribution (see Fig.~\ref{f:omegab_signals}) extracted to a signal of $17.8 \pm 4.8 (stat)$ events, with an estimated mass of $6165 \pm 10 (stat) \pm 13 (syst) MeV$.

\subsection{The CDF observation}
\vskip .3cm

CDF searched for the same final state of $\Omega_b$ decays employed by D0, but used a dataset over three times as large, corresponding to an integrated luminosity of $4.2 fb^{-1}$. Instead of optimizing the sensitivity to the $\Omega_b$ signal with multivariate techniques, CDF opted for a conservative selection aimed at straight, simple requirements, which allowed a comparison of the reconstruction of the searched baryon with that of B meson signals with similar decay topologies, notably $B^0 \to J/\psi K^*(892)^0 \to \mu^+ \mu^- K^+ \pi^-$ and $B^0 \to J/\psi K^0_s \to \mu^+ \mu^- \pi^+ \pi^-$. An excellent (12 MeV) mass resolution for the $\Omega^-$ candidates was obtained by imposing that the reconstructed flight direction of $\Lambda$ particles intersect with the helix of the negative kaon. Backgrounds were reduced by selecting candidates with decay lenghts exceeding 1 cm; silicon hits belonging to particle tracks were used when available, to improve the mass resolution of the decay products and further suppress backgrounds.

Mass and yield of the $\Omega_b$ baryon were extracted with a likelihood fit to the subset of candidates having a reconstructed decay length exceeding $100 \mu m$, a requirement predicted to reduce backgrounds significantly. CDF thus measured a mass of $6054.4\pm6.8(stat)\pm 0.1 (syst) MeV$ and a yield of $12\pm 4$ events. An alternative fit using both mass and lifetime information together allowed to confirm the signal, extracting a signal of $16^{+6}_{-4}$ events and a probability of the null hypothesis $P=4\times 10^{-8}$, which corresponds to a significance of 5.5 standard deviations. Figure~\ref{f:omegab_signals} shows the mass distribution of the candidates extracted by CDF, for three increasing values of the selection requirement on the $\Omega_b$ decay length. Similarly extracted signals of $\Lambda_b$ and $\Xi_b$ baryons allowed a measurement of the production rate of $\Omega_b$ particles relative to that of the other baryons. Mass, lifetime, and production rate of the signal obtained by CDF agree with theoretical estimates~\cite{theor_omegab}.

\subsection {An oversight in the D0 significance calculation}
\vskip .3cm

The paper published by D0 on their $\Omega_b$ observation~\cite{d0_omegab} may leave the reader with a doubt on the exact procedure by which that number is evaluated. Here is the relevant text: \par

{\em ``To assess the significance of the excess, we first determine the likelihood $L_{(s+b)}$ of the signal plus background fit above and then repeat the fit with only the background contribution to find a new likelihood $L_b$. The logarithmic likelihood ratio $\sqrt{ 2 ln(L_{(s+b)}/L_b)}$ yields a statistical significance of $5.4 \sigma$, equivalent to a probability of $6.7 \times 10^{-8}$ that the background could fluctuate with a significance equal to or greater than what is observed.''}

Therefore D0 used a standard "log-likelihood approach" to estimate the significance of the peak they find in the reconstructed mass distribution. The two concurrent hypotheses were compared using the ratio of their likelihoods: a null hypothesis, according to which the data is only produced by background processes and has a reconstructed mass distributed with a uniform probability density function (PDF); and the alternate hypothesis that the data contains a Gaussian signal on top of a uniform background. The width of such a signal was assumed known, since the $\Omega_b$ decays weakly and thus the Gaussian shape only reflects the known experimental resolution in the momenta of final state particles, which can be modeled by Monte Carlo simulations; the $\Omega_b$ mass, however, was assumed {\em a priori} unknown, and so was the production rate. 
\begin{center}
\begin{figure}
\centerline{\epsfig{file=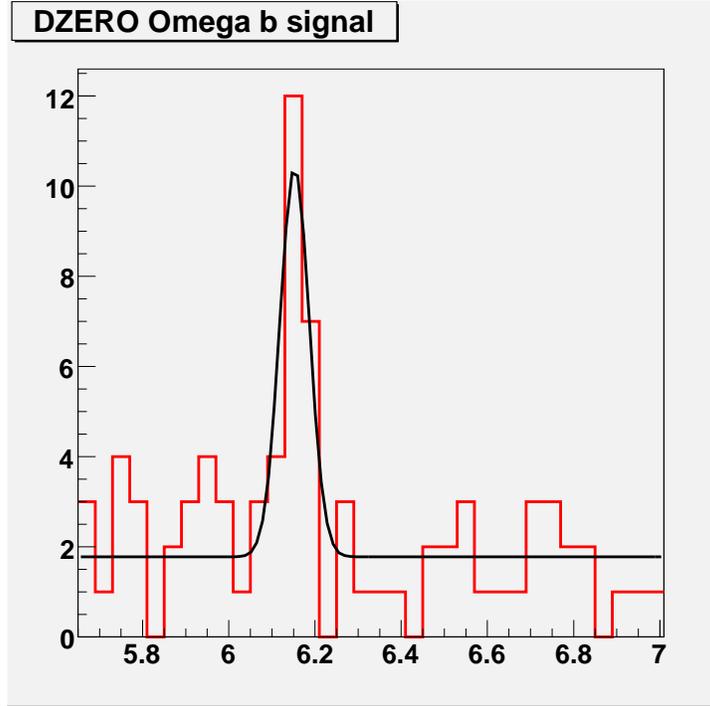, bb = 0 0 600 600, width=10cm, angle=-90, clip=}}
\caption{\em Result of a binned likelihood fit to the mass histogram of the 79 D0 $\Omega_b$ candidates. The fit returns a signal of 18.4 events. }
\label{f:myd0sig}
\end{figure}
\end{center}
The histogram in Fig.~\ref{f:omegab_signals} (left) has only 34 bins, and the number of entries in each of them can be clearly read out on the vertical axis. It is therefore quite easy to recreate an identical histogram, fit it with a uniform PDF, note the likelihood returned by the fitter, and repeat the fit for the alternate hypothesis, which includes a Gaussian of fixed width and varying mean and normalization. Figure~\ref{f:myd0sig} shows the result of this exercise, in the case of the fit to the signal-plus-background hypothesis: the result matches well with the original one. The parameters of the fit function are not identical, due to the fact that the original fit performed by D0 is unbinned~\footnote{Individual mass values of the 79 D0 candidates are unavailable.}; however, given that the bin width is comparable in size with the experimental mass resolution, those differences do not affect the conclusions which may be drawn from the result. If we liken the value of $-2 \Delta( \log L)$ to a $\chi^2$ {\em with one degree of freedom}, we may compute the probability of the null hypothesis as $P(\chi^2,1)$; from the probability we arrive at a significance using the formula\\

\begin{center}
{\large
$ N = \sqrt{2} \times Erf^{-1} [1-P(\chi^2,1)]$
}
\end{center}
\normalsize

\noindent where $Erf^{-1}$ is the inverse of the error function. By this recipe we get the numbers in the middle column of the table below\footnote{One caveat is that the delta-log-likelihood will in general only distribute as a $\chi^2$ function only in the asymptotic limit of large number of entries per bin, which is not a regime which applies here.} : \\


\begin{center}
\begin{tabular}{l|c|c}
	                                       & This study           & Ref.~\cite{d0_omegab}\\
\hline 
	Likelihood of null hypothesis          & 70.03                & not quoted \\
	Likelihood of the alternate hypothesis & 55.38                & not quoted \\
	Probability of the null hypothesis     & $6.2 \times 10^{-8}$ & $P=6.7 \times 10^{-8}$ \\
	Significance of the observed signal    & 5.41 st. dev.        & 5.4 st. dev. \\
\hline
\end{tabular}
\end{center}

Despite the approximation of using a binned likelihood to fit the mass histogram, the results of~\cite{d0_omegab} are nicely confirmed. Unfortunately, the above conversion of the delta-log-likelihood value into a probability, and therefore also the significance quoted by D0, is erroneous. In fact, when one fits the data with the alternate hypothesis, one is adding {\em two} free parameters to it: not just the normalization of the Gaussian signal, but also its free-floating mass. On this point the D0 article specifies that {\em ``We fix the Gaussian width to 0.034 GeV, the width of the MC $\Omega_b$ signal''}. It is reasonable to assume that if the fitting procedure had also entailed fixing either the $\Omega_b$ mass or yield (both of which are {\em a priori} unknown) in the fitting function for the alternate hypothesis, the paper would have reported it. Thus, D0 appears to be throwing in {\em two} extra degrees of freedom to find the $\Omega_b$ signal, and a correct calculation of the significance must account for that, because the factor $-2 \Delta (\log L)$ will distribute like a $\chi^2$ with two, and not just one, degrees of freedom. By taking this into account, the following results are obtained with the likelihood values quoted above:\par

\begin{itemize}
    \item Probability of the null hypothesis: $P=4.3 \times 10^{-7}$;
    \item Significance of the observed signal: 5.05 standard deviations.
\end{itemize}

\noindent
A $5.05 \sigma$ signal is seven times more probable to result from a statistical fluctuation than a $5.4 \sigma$ one, so the detail is worth noticing. This inaccuracy in the D0 publication alone does not of course disprove the repoted observation, although it decreases the strength of the result. A pseudoexperiment-based approach may be used to further check the significance. 

\subsection{A check of the D0 signal significance with pseudoexperiments \label{s:d0significance}}
\vskip .3cm

A pseudoexperiment mimicking signal extraction from the D0 $\Omega_b$ candidates consists in filling a 34-bin mass distribution of range equal to the $\Omega_b$ histogram shown in Fig.~\ref{f:myd0sig} with 79 entries, sampled according to a uniform PDF, then trying to fit a Gaussian signal somewhere in the spectrum~\footnote{ Although it might be argued that a preferable procedure consists in varying the number of entries for each pseudoexperiment according to a Poisson distribution, we avoid this simple modification because what we mean to test is the calculation of the significance of the actual histogram used by D0. This kind of {\em conditioning}~\cite{cousins,reid} is generally accepted by statisticians. } on top of a constant background. 

The straightforward procedure includes two steps per each pseudo-histogram~\footnote{
We note here that there are a unavoidable set of software implementation details and arbitrary choices that potentially affect the results of a test with pseudoexperiments. Among the former we may quote the choice of random-number generator employed for the construction of pseudo-data templates; we use TRandom3~\cite{root}, following the recommendation of the developers of the {\em root} package. Among the latter are the initial value of fit parameters modeling the alternate hypothesis, their allowed range ({\em e.g.} if the normalization of the Gaussian distribution is enforced to be positive), and the procedure by which their full scan is enforced. Pseudoexperiment results are thus not immune from systematic effects; nevertheless, if very well-defined questions are posed the answers are typically well-reproducible.  }:
a fit to a uniform PDF testing the null hypothesis, and then a fit of the same histogram adding to the uniform PDF the two degrees of freedom of a Gaussian signal, of width fixed to the experimental mass resolution quoted by D0 (the alternate hypothesis). The difference between the likelihood of the two fits can finally be converted in a significance, using the recipe described in the previous section. From repeated trials of this procedure, a picture can be obtained of the probability distribution of the number of signal events that may be fit in the absence of a signal, and the corresponding computed significance. One may thus infer by a ratio between successes and total trials whether the mass bump found by D0 is something that happens by chance only 6.7 times every hundred-millions (as D0 claims), or rather 4.3 times every ten millions (as obtained by using two degrees of freedom in the calculation), or more frequently so; the distribution of significances from the $-2 \Delta(\log L)$ values is additional information that may be used to check the frequency ratio. 

\begin{center}
\begin{figure}
\begin{minipage}{0.49\linewidth}
\centerline{\epsfig{file=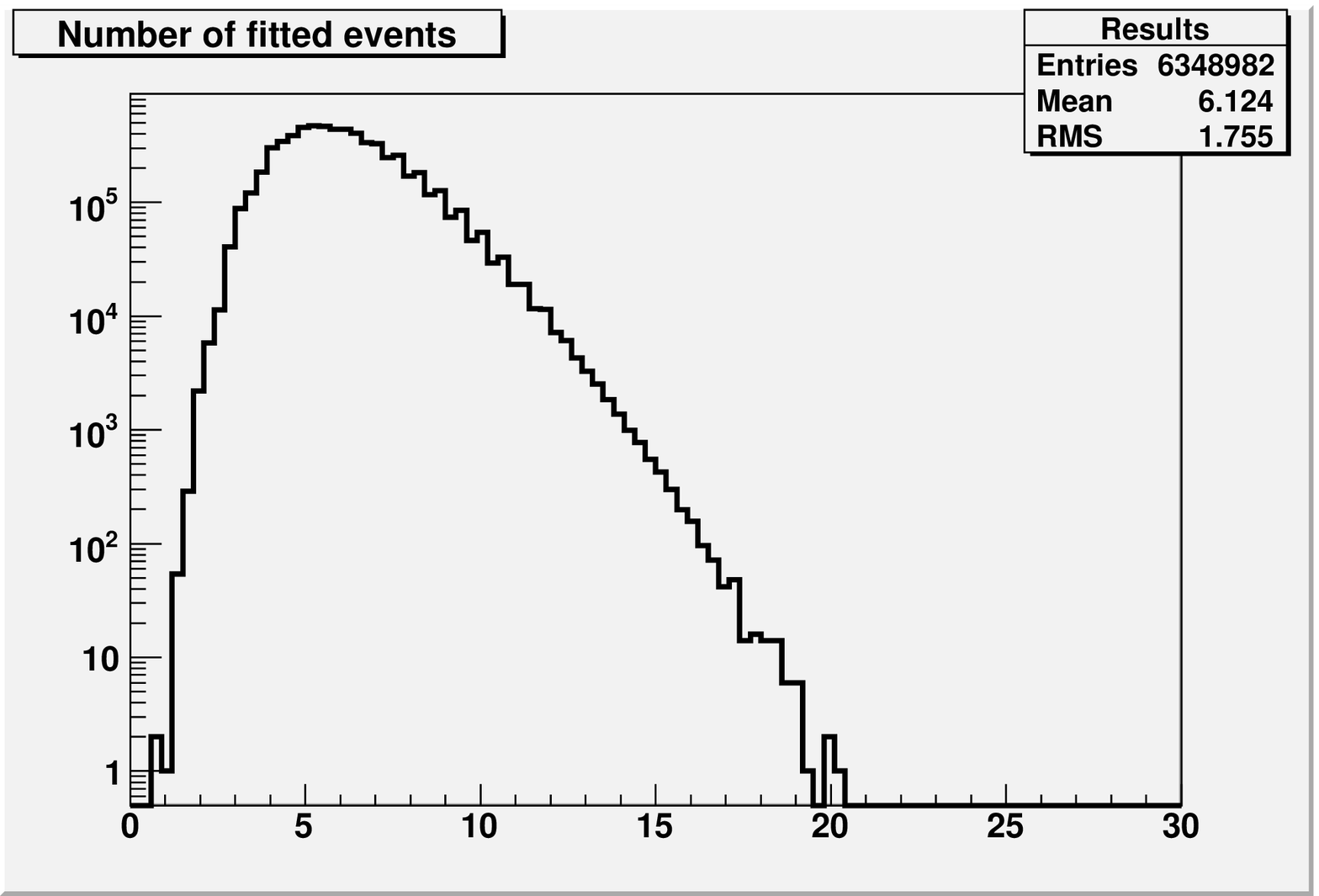, bb=0 0 600 600, width=8cm, angle=0, clip=}}
\end{minipage}
\begin{minipage}{0.49\linewidth}
\centerline{\epsfig{file=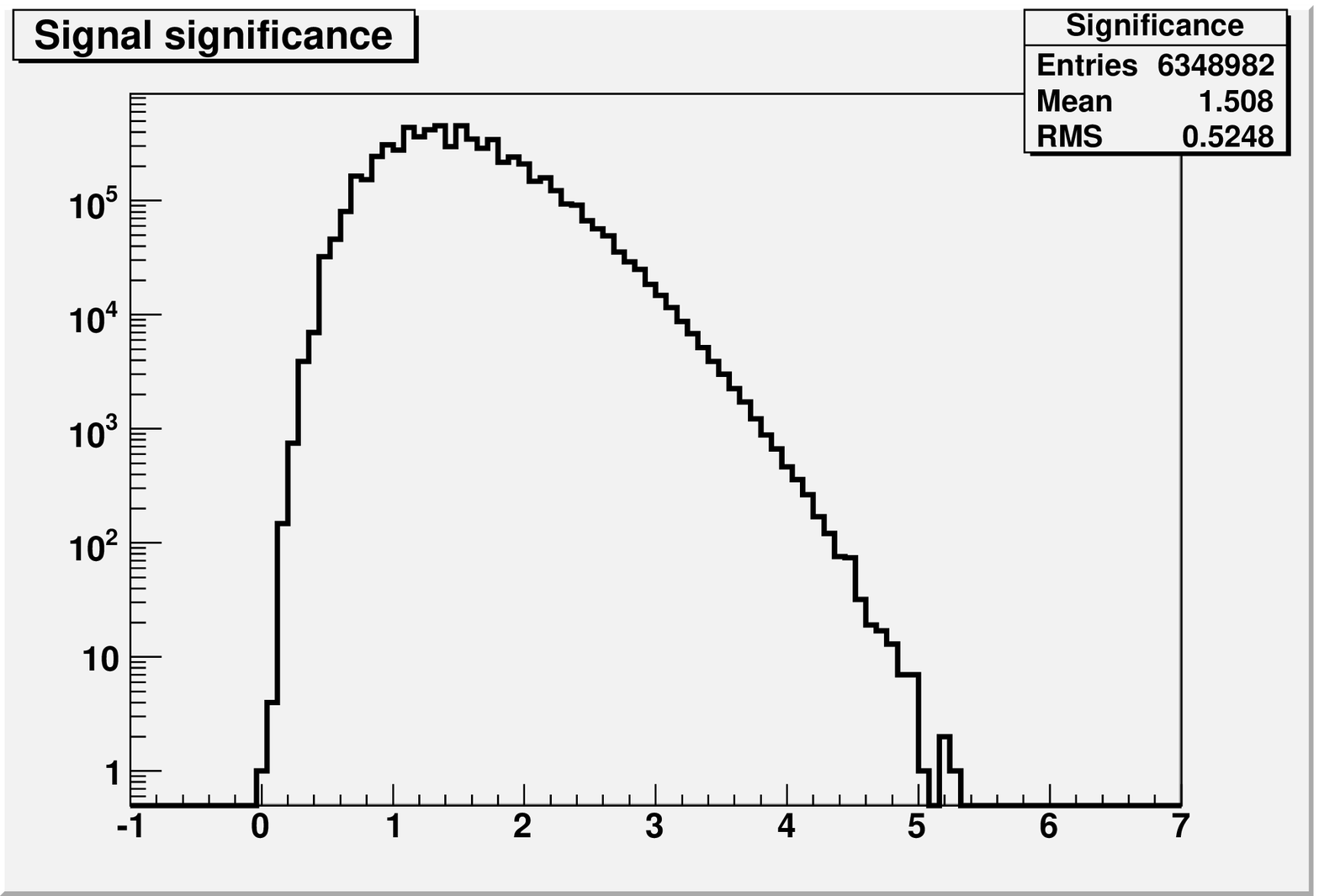, bb= 0 0 600 600, width=8cm, angle=0, clip=}}
\end{minipage}
\caption{\em left: Number of signal events in the Gaussian (of width equal to the D0 mass resolution) fit in pseudo-histograms containing 79 events. Right: Significance of D0-like pseudoexperiments, computed using delta-log-likelihood values. The wiggling of the distributions reflects the discreteness of the distributions created by the pseudoexperiments.}
\label{f:d0_pe_nevts}
\end{figure}
\end{center}

In a 6,348,982 pseudoexperiments run, 26 histograms resulted in a signal of just above 18.4 events\footnote{The binned-likelihood fit to the D0 mass histogram returns 0.6 more events than the unbinned fit quoted by D0; of course, the results of binned-likelihood fits to pseudo-histograms must be compared to the result on real data obtained with the same method. }, with a corresponding probability of $P_{null}=4.1 \pm 0.8\times 10^{-6}$, equivalent to a significance of 4.61 standard deviations. The distribution of the number of events returned by the fit and corresponding significance are shown in Fig.~\ref{f:d0_pe_nevts}.

\begin{center}
\begin{figure}
\begin{minipage}{0.49\linewidth}
\centerline{\epsfig{file=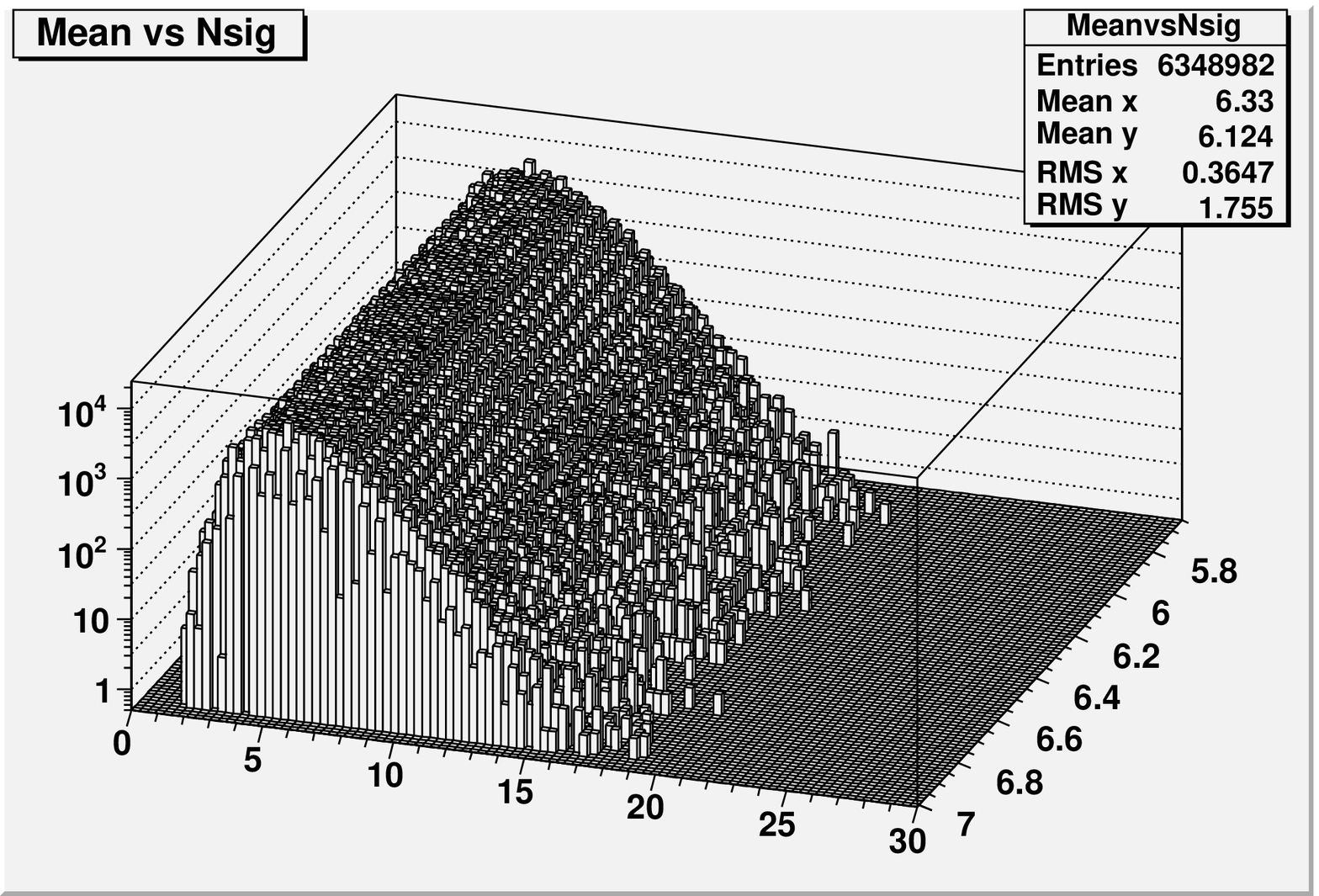, bb=0 0 600 600, width=8cm, angle=0, clip=}}
\end{minipage}
\begin{minipage}{0.49\linewidth}
\centerline{\epsfig{file=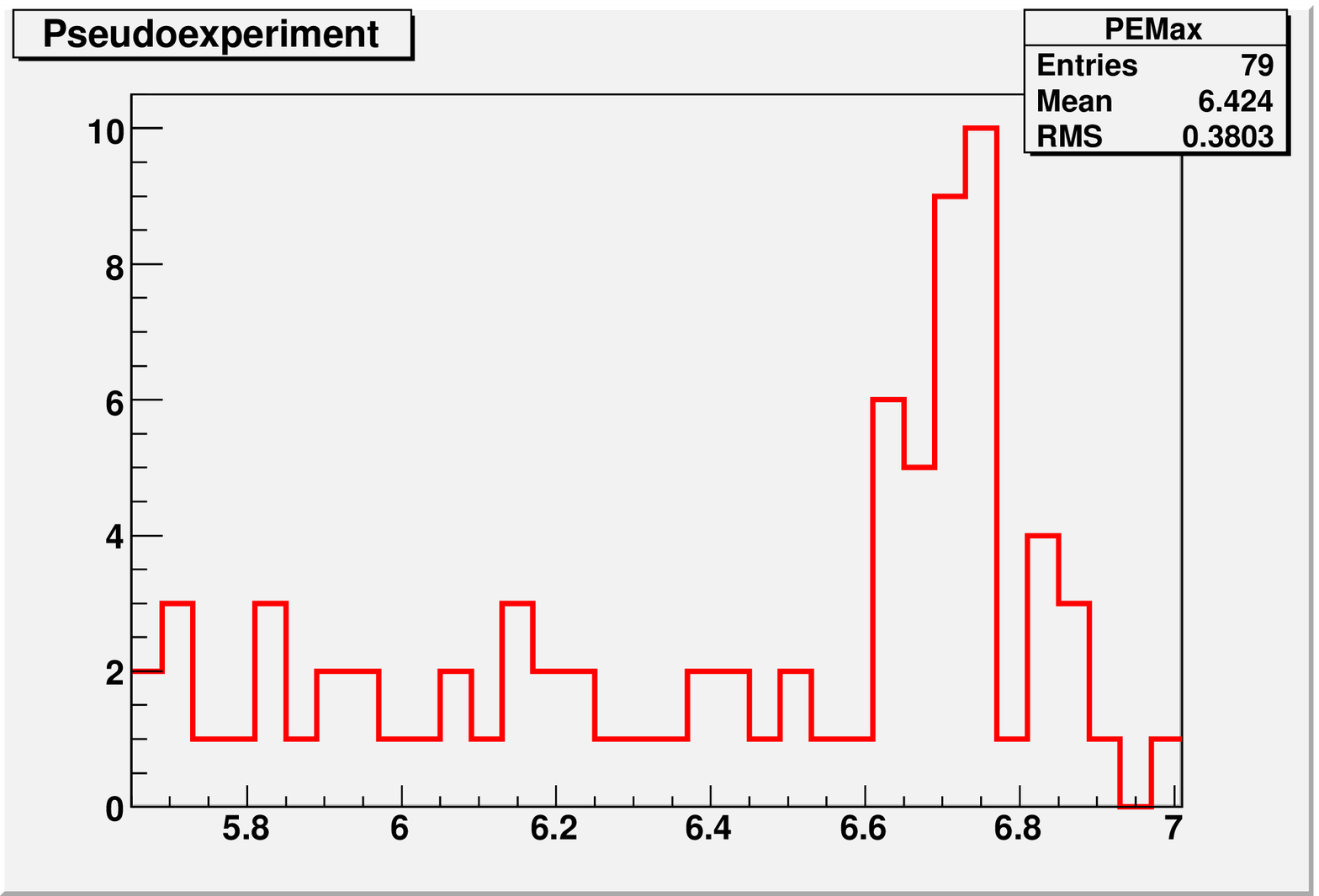, bb= 0 0 600 600, width=8cm, angle=0, clip=}}
\end{minipage}
\caption{\em Left: number of fit events versus fit mass in the background-only pseudoexperiments mimicking the experimental data of D0. Right: the pseudo-histogram which mimics the largest signal in the set of over 6.3 millions. }
\label{f:d0pe_2}
\end{figure}
\end{center}

It is instructive to also examine the additional distributions shown in Fig.~\ref{f:d0pe_2}: there one clearly observes that the Gaussian degrees of freedom are fully exploited by the fit, which scans the mass values in search for the most profitable way to use the two extra degrees of freedom (mass and normalization of the Gaussian) to increase the likelihood. But the two degrees of freedom only account for {\em local} adjustments of the global likelihood, while the freedom of the fit (or rather, the experimenter) to set his or her attention on a fluctuation anywhere in the distribution amounts to a further derating of the actual significance: this is the well-known ``look-elsewhere effect''. Since the signal has a width of size similar to the bin width, and will thus typically involve $\sim 3$ adjacent bins simultaneously fluctuating upwards, one expects that the true probability of a signal occurring in a non-a-priori-specified spot is larger than the probability to observe it in any specific spot by a {\em trials factor} of the same order of magnitude of $N_{bins}/3$, which is roughly the number of independent regions where a signal may be sought. This is well borne by the ratio between the probability computed by the pseudoexperiments $P_{null}$ and the probability computed with the likelihood difference $P_{\Delta}$\footnote{
     In order to take in account the effect of the hand and the eye of the experimenter, who is called to provide the fit with a starting value of all free parameters, the fits to the pseudo-histograms are performed by choosing a sufficient number of points along the $x$ axis (10 in our case) as starting mass values; the fit returning the maximum likelihood among the set is then considered the one that would be picked in a real experiment. Fit results improve with this procedure, which invites the minimization routine to scan fully the parameter space in order to search for the best fluctuation. As far as the normalization of the Gaussian is concerned, fit results depend much less strongly on its starting value; zero events were used for the tests described here, but the normalization was constrained to be positive in the fit. }: 

\begin{center}
{\large
$P_{null}/P_{\Delta} = \frac {4.1 \pm 0.8 \times 10^{-6}}{4.3 \times 10^{-7}}=9.5\pm1.9 \sim 10.33 = N_{bins}/3$. 
}
\end{center}
\normalsize

Our results are not in contrast with what is reported in a document answering frequently-asked questions on the $\Omega_b$ observation, which was made available on May 27, 2009 in the D0 collaboration web site~\cite{d0faq}: \par

{\em What we report is purely the statistical significance based on the ratio of likelihoods under the signal-plus-background and background-only hypotheses. Therefore, no systematic uncertainties are included, although we have verified that, after all systematic variations on the analysis, the significance always remains above five standard deviations. Our estimate of the significance does also not include a trials factor. We believe this is not necessary since we have a specific final state (with well-known mass resolution) and a fairly narrow mass window (5.6-7.0 GeV) where we are searching for this particle. [...]}

The statement above, according to which the inclusion of a trials factor is unnecessary, given that a narrow mass window is tested, is surprising. Indeed, the inclusion of a trials factor of about a factor of 10 is necessary, as proven above.


It is instructive to inspect the pseudo-histogram with the largest signal among the generated ones in Fig.~\ref{f:d0pe_2}. Once in a few million cases, a real beautiful peak does appear by chance!

\section{A discrepancy and its possible causes \label{s:mass_riddle}}
\vskip .5cm

Regardless of the notes made in the previous section about the real significance of the D0 signal, each of the two observations of the $\Omega_b$ resonance appears to provide, at first sight, convincing evidence of the existence of this heavy baryon: in both cases one observes a quite distinctive decay chain: one not very different from that which convinced physicists of the existence of the $\Omega^-$ (sss) baryon in 1964, based on one single striking image obtained by the 80-inch bubble chamber at Brookhaven. However, the two results constitute an embarassing problem if taken together. The CDF paper puts the matter very bluntly in the introduction:

{\em"In this paper, we report the observation of an additional heavy baryon and the measurement of its mass, lifetime, and relative production rate compared to the $\Lambda_b$ production. The decay properties of this state are consistent with the weak decay of a b-baryon. We interpret our result as the observation of the $\Omega_b$ baryon ($|ssb>$). Observation of this baryon has been previously reported[6], however, the analysis presented here measures a mass of the $\Omega_b$ to be significantly lower than ref.[6]"}.

Reference [6] above is, of course, the D0 paper, which we have discussed in section 2. Let us place side-by-side the two $\Omega_b$ mass determinations: \par

	\begin{itemize}
	\item     $M_{D0} = 6165 \pm 10 \pm 13 MeV = 6165 \pm 16.4 MeV$;
    	\item     $M_{CDF} = 6054.4 \pm 6.8 \pm 0.9 MeV = 6054.4 \pm 6.9 MeV$.
	\end{itemize}

\noindent where statistical and systematic uncertainties quoted by each experiment have been added in quadrature for the sake of comparing the nominal total error of the two determinations.

One should immediately note a few things from the numbers above. One: the CDF measurement has an error bar 2.4 times smaller than the D0 measurement. Two: the D0 measurement has a systematic uncertainty which is twice as big as the total uncertainty of the CDF measurement. Three: the systematic part of the uncertainty in the CDF measurement ($0.9 MeV$) is virtually irrelevant --a by-product of the performant charged particle tracking of CDF, its careful calibration, and the analysis method chosen in the search of $\Omega_b$ candidates. Four: the two mass determinations differ by $110.6 MeV$. If we were to fully trust the quoted CDF and D0 uncertainties, we would have to conclude that the two experiments have measured two distinct particles! In fact, their "difference", in units of total uncertainty, is of 6.2 standard deviations if one adds all uncertainties in quadrature; more conservative recipes do not change the picture appreciably. Given the quoted significances of the two signals stand in the whereabouts of 5 standard deviations each (4.6-sigma as estimated in section~\ref{s:d0significance} for the D0 signal, 5.5-sigma for CDF\footnote 
{As noted above, the insufficient information provided in the CDF publication prevents a check of the significance they claim for their $\Omega_b$ signal, which is obtained from a two-dimensional fit to mass and lifetime together; we should expect that a proper accounting of the ``look-elsewhere effect'' might decrease the CDF significance by a similar amount as what we observe for the D0 signal; Ref.~\cite{cdf_omegab} however notes, referring to its mass-only fit, that {\em ``This calculation was checked by a second technique, which used a simulation to estimate the probability for a pure background sample to produce the observed signal anywhere within a $400 MeV/c^2$ range. The simulation result confirmed the significance obtained by the ratio-of-likelihoods test''}. Being unable to verify this statement, we have to rely on it; we expect that a larger search window would have derated the quoted significance by a few decimal points.}
), {\em it looks more likely that these be two distinct particles, rather than either of them be simply a fluctuation of backgrounds}. The matter requires some further investigation, which is offered below.

\subsection{Hypotheses for the discrepancy \label{s:twohyp}}
\vskip .5cm

The two mass measurements disagree by more than six standard deviations, and they both possess an observation-level significance, if barely so. The hypothesis that the two signals represent different particles cannot be discarded {\em a priori}, but  this is at the very least problematic: one should then explain why each experiment only sees one of the two states. An approach to the problem can be to try and determine how likely it is that only one of the two results is correct, and the other is wrong. 

Let us admit for the sake of argument that the CDF result is correct: you may then take the $\Omega_b$ mass and production rate as measured by CDF, and plug these numbers in a pseudoexperiment generation, to determine what would D0 be expected see in their data under such conditions, and how likely it is that they would find a significant signal at a $110.6 MeV$ or larger distance from $M_{CDF}$.

Some additional input is needed in order to carry out this exercise: specifically, we need to determine the rate of $\Omega_b$ events which would be accepted by the D0 selection under the hypothesis that the production rate of that particle is the one computed by CDF. In Ref.\cite{cdf_omegab} a rate comparison is indeed offered:\par

{\em ``The relative rate measurement presented in Ref.[6] is $\frac{f(b \to \Omega_b^-){\cal{B}}(\Omega_b^- \to J/\psi \Omega^-)}{f(b \to \Xi_b^-){\cal{B}}(\Xi_b^- \to J/\psi \Xi^-)}=0.80 \pm0.32(stat)^{+0.14}_{-0.22} (syst)$ where $f(b \to \Omega_b^-)$ and $f(b \to \Xi_b^-)$ are the fractions of b quarks that hadronize to $\Omega_b^-$ and $\Xi_b^-$. The equivalent quantity taken from the present analysis is $\frac{\sigma(\Omega_b^-){\cal{B}}(\Omega_b^- \to J/\psi \Omega^-)}{\sigma(\Xi_b^-){\cal{B}}(\Xi_b^- \to J/\psi \Xi^-)}=0.27 \pm0.12(stat) \pm 0.01 (syst)$. Neither measurement is very precise, since a ratio is taken of two small samples. Nevertheless, this analysis indicates a rate of $\Omega_b^-$ production substantially lower than Ref.[6]''}.

Noting that the D0 estimate of the rate fraction of $\Omega_b$ and $\Xi_b$ baryons already includes the statistical uncertainty in the number of $\Omega_b$ candidates, the number of events to generate in a D0-like pseudoexperiment may be calculated as \par

\begin{center}
\large{
$N^{exp}_{D0 | CDF} = 17.8 \times \frac {0.27\pm0.12\pm 0.01}{0.80\pm0.32^{+0.14}_{-0.22}} = 6.0^{+3.95}_{-3.74}$},
\end{center}
\normalsize

where, for simplicity, statistical and systematical uncertainties of each rate determination have been added in quadrature, and Gaussian distributions have been assumed. For later use, the corresponding number $N^{exp}_{CDF|D0}$, referring to the $12\pm 4$ events extracted by the mass-only fit of the distribution shown in the bottom histogram of Fig.~\ref{f:omegab_signals}, is also given below~\footnote {Since the mass-only fit by CDF refers to the sample with decay length above $100 \mu m$, while the relative rate is obtained by relaxing that requirement and performing a combined mass-lifetime fit, in principle a part of the statistical error on the absolute rate should be retained in the error propagation; because of insufficient information this effect is neglected. We also ignore the possible difference between the rate of events in the loose and tight selections. }:

\begin{center}
\large{
$N^{exp}_{CDF | D0} = 12 \times \frac {0.80\pm0.32^{+0.14}_{-0.22}}{0.27\pm0.12\pm 0.01} = 35.6^{+22.2}_{-23.4}$}.
\end{center}
\normalsize 

Mass distributions of 79 events can then be generated by taking $6.0^{+3.95}_{-3.74}$ entries (allowing for variations within uncertainties) from a Gaussian distribution of mass equal to $M_{CDF}=6054.4 \pm 6.9 MeV$ and width equal to $34 MeV$ (D0's experimental mass resolution), and the remaining ones from a uniform background PDF. The resulting histogram can be fit {\em a' la D0}, as done in the previous section. Under normal conditions, the fit will return what was given in input: a signal of about $N_{fit}=6$ events, sitting at $M_{fit} \sim M_{CDF}$ (give or take fifteen MeV or so). One might wonder, however, whether an upward fluctuation of the number of generated signal events might occasionally conspire with a weird background fluctuation occurring at masses just above or below $6054 MeV$ to create a bump yielding a larger number of $\Omega_b$ decays, at a mass significantly different than $M_{CDF}$. As discussed in section~\ref{s:d0significance}, a binned likelihood fit to the distribution observed by the D0 collaboration results in a signal of 18.4 events. How often does this happen for $|M_{fit}-M_{CDF}|>110.6 MeV$ in the situation just hypothesized ? Well, not quite often: in a million-pseudoexperiments run, no such occurrences are observed.

\begin{center}
\begin{figure}
\begin{minipage}{0.49\linewidth}
\centerline{\epsfig{file=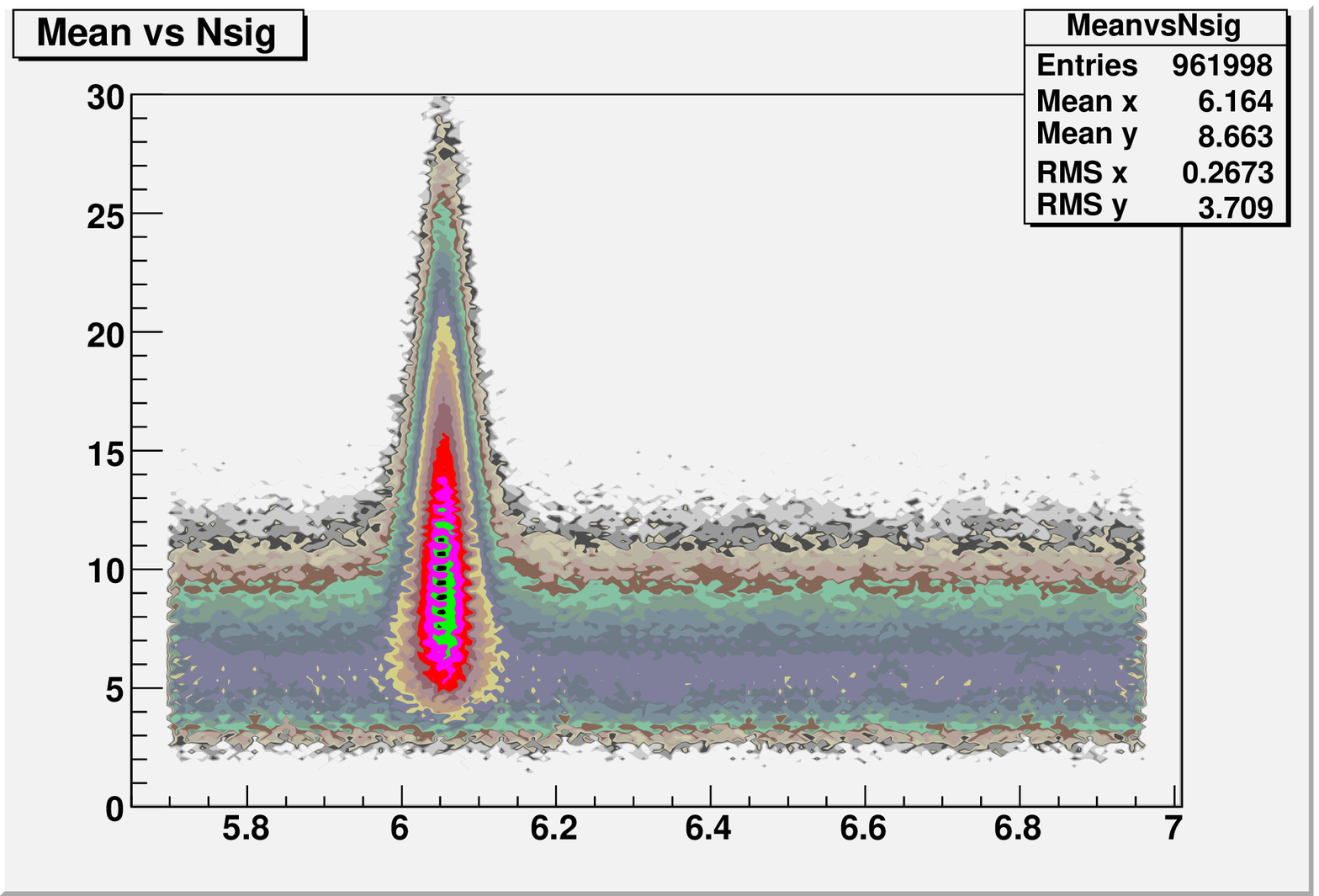, width=7.5cm, angle=0, clip=}}
\end{minipage}
\begin{minipage}{0.49\linewidth}
\centerline{\epsfig{file=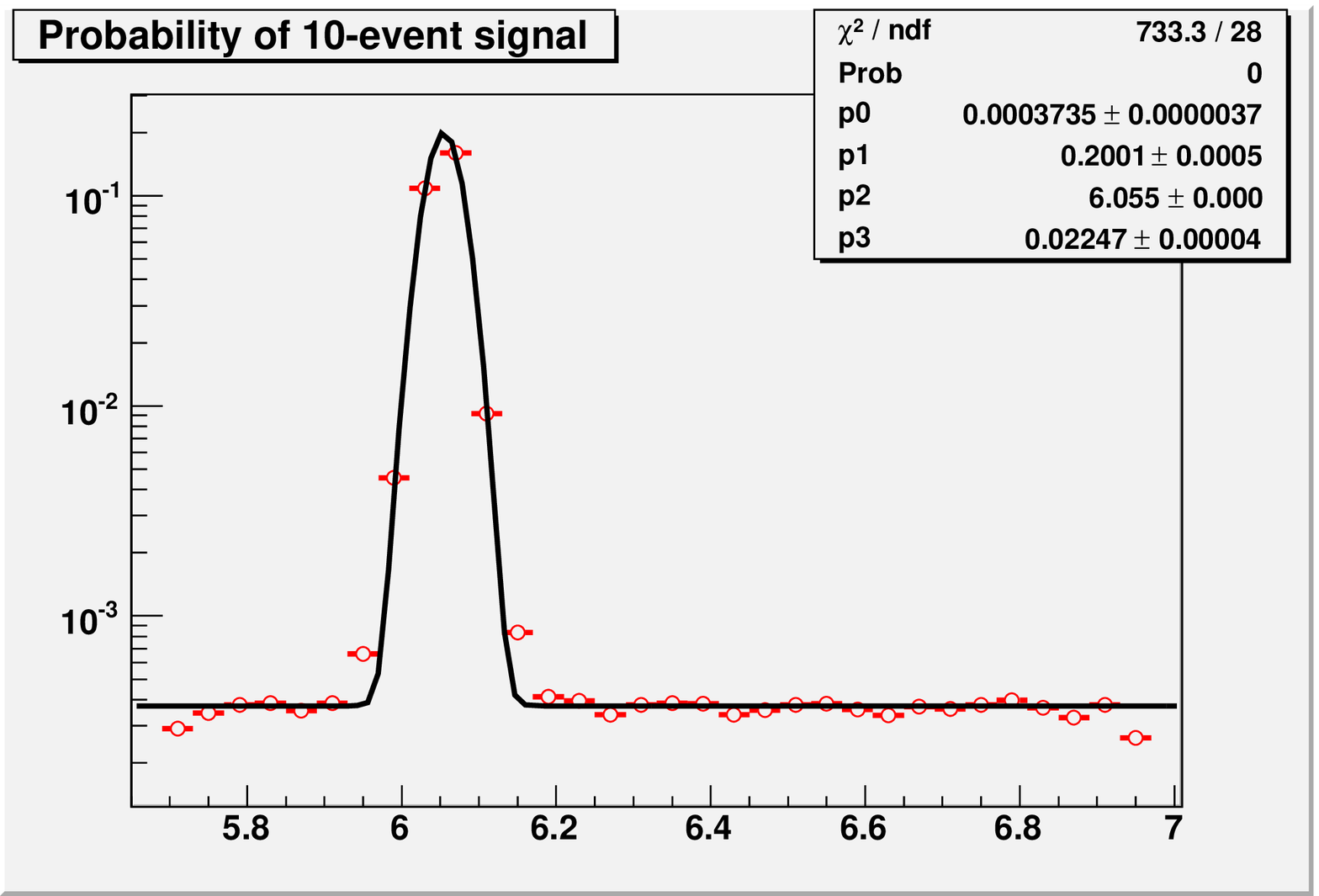,  width=7.5cm, angle=0, clip=}}
\end{minipage}
\caption{\em Results of 961,998 binned-likelihood fits to 79-event pseudo-histograms generated with a CDF-like $\Omega_b$ signal of normalization chosen as explained in the text. Left: number of events returned by the fit as a function of the fit mass; right: relative fraction of pseudoexperiments returning a signal of $N_{fit}>10$ events. Note the outlier bins above the fit line on each side of the Gaussian.}
\label{f:D0givenCDF}
\end{figure}
\end{center}

In the graphs of Fig.~\ref{f:D0givenCDF} one can see that a large fraction of the million pseudoexperiments return the correct answer: mass and normalization close to the generated ones. It may also be noted that the fit is not prevented from sometimes using the two Gaussian degrees of freedom to model a different fluctuation happening elsewhere in the spectrum: that is the cause of the band spanning the whole mass range in the left graph of Fig.~\ref{f:D0givenCDF}. The band is centered at $N_{fit} \sim 6$ events by pure chance; it is a feature which depends on to the number of generated data in the fitted histogram and the width of the signal which is sought. 

From the test one also observes that there are two effects at work in determining the likelihood of fits returning a large signal at a significantly displaced mass value from the generated one. The first is the fact that the presence of a 6-event signal at a mass $M_{CDF}$ reduces the number of events distributed uniformly in the mass spectrum, and conseuqently the likelihood of a large Poisson fluctuation of the background alone; the second is the possibility of a ``spill-over'' of the generated signal, due to the combination of fluctuations and binning effects~\footnote{ It is important to reiterate here that since these tests are based on a binned-likelihood fit, the conclusions that may be drawn from them are only approximately valid for the experimental situation of D0.}. The former effect needs no commentary; the latter is maybe better appreciated by examining the right panel in Fig.~\ref{f:D0givenCDF}. There, the fraction of pseudoexperiments returning a signal of 10 or more events is displayed as a function of $M_{fit}$~\footnote { A signal of 10 or more events rather than 18.4 has been chosen to illustrate qualitatively the effect because the statistics of pseudoexperiments returning 18.4 or more events is too scarce.}. We observe a probability of $3.7 \times 10^{-4}$ that a fluctuation of 10 or more events is picked up by the fit away from $M_{CDF}$, where the signal is actually generated. However, even in the cases when the fit does converge to masses in the vicinity of the generated mass (which are the vast majority), we do see a significant non-Gaussian tail in $M_{fit}$. This tail is wide enough to influence the probability of fitting masses at $|M_{fit}-M_{CDF}|>110.6 MeV$, at least for the case of $N_{fit}>10$ signal events. 

We conclude that the observed D0 signal can hardly be attributed to a ``statistical leak'', a spill-over of a nearby peak, if the CDF mass {\em and rate} measurements are correct. The issue will be revisited when we modify some of the assumptions, in section~\ref{s:twosf}. There, we will see that the leak may indeed have an impact in formulating a hypothesis for the CDF/D0 controversy on the $\Omega_b$ baryon.

\subsection{The first hypothesis: underestimated mass systematics \label{s:d0massbias}}
\vskip .3cm

If the Tevatron experiments had measured the $\Omega_b$ mass value with larger error bars there would be no controversy, but just a mild disagreement in the measured rate of $\Omega_b$ production. The relevant uncertainties we need to put under scrutiny are of course the systematical ones; we observe that the D0 mass systematic uncertainty ($(\delta M)^{syst}_{D0} = 13 MeV$) totally outweighs the CDF one ($(\delta M)^{syst}_{CDF} = 0.9 MeV$). Therefore, one plausible hypothesis of the source of the CDF/D0 discrepancy lies in non-well-controlled systematical uncertainties in the mass determination produced by the D0 collaboration. In the already cited frequently-asked-questions document~\cite{d0faq}, they note: \par

{\em We have adopoted an approach where the systematic uncertainty on the mass is estimated by comparing the measured mass value after performing small variations to the analysis (e.g. different selection criteria). At this level of statistics, this introduces a significant statistical component to this systematic uncertainty which is expected to be reduced in the future with larger data sets or simply by performing a more refined evaluation of the systematic uncertainty via large MC samples[...]}.

While waiting for those studies, we may test the hypothesis that the D0 mass systematics are underestimated by inflating them with a multiplicative scale factor $k$, and using that value in pseudoexperiments, to assess by what factor would the D0 mass systematics need to be increased in order to obtain a reasonable probability that they observed 18.4\footnote{As already noted, to be consistent we take our binned-likelihood result as a reference value for the D0 signal.} or more events at a mass of $M_{D0}=6165 MeV$ or further away from $M_{CDF}$. 

\begin{center}
\begin{figure}
\centerline{\epsfig{file=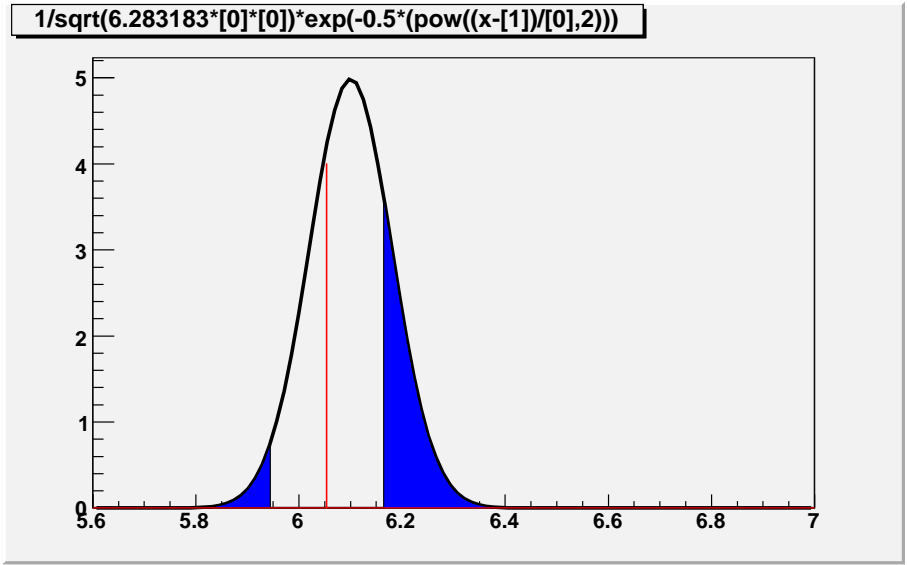,  width=12cm, angle=0, clip=}}
\caption{\em Graphical explanation of the method used to compute the probability of a mass measurement at least as distant from the CDF measurement as the one obtained by D0. The red line is at $M_{CDF}=6054.4 MeV$; the Gaussian has a mean equal to the fit mass of a pseudoexperiment finding a signal exceeding 18.4 events, and a width equal to the one expected if a large scale factor $k$ (exaggerated for display purpose) is hypothesized for the systematical uncertainty in the D0 mass measurement. By integrating the blue area of the Gaussian, we obtain an estimate of the probability that a value as discrepant as the D0 one was observed.}
\label{f:gauss}
\end{figure}
\end{center}

Since we mean to test an ensemble of situations, each of which has a tiny probability of occurrence, it is too CPU-time consuming to rely on the usual pseudoexperiment calculation of a probability as a number of successes divided by number of trials. One may circumvent this problem by performing a different but asymptotically equivalent calculation, which is graphically illustrated in Fig.~\ref{f:gauss}. For each hypothesis on the scale factor $k$ we may perform pseudoexperiments in which pseudo-histograms contain 79 events, $6.0^{+3.95}_{-3.74}$ of which are taken to be signal events with a mass $M_{CDF}=6054.4 \pm 6.9 MeV$. This time, for each fit returning 18.4 or more signal events we integrate in the range $\Lambda = [-\infty,2 M_{CDF}-M_{D0}] \cup [M_{D0},\infty]$ a Gaussian function centered at $M_{fit}$, of width equal to $\sigma(k) = \sqrt{10^2+k^2 \times 13^2} MeV$, normalizing by the number of trials:

\large
\begin{center}
$P(k) = \frac {\sum_{(N>18.4)}{\frac{1}{\sqrt{2 \sigma(k)}} \int_{\Lambda} e^{-\frac{(M_{fit}-x)^2}{2 \sigma(k)^2}}} dx}{N_{trials}}$.
\end{center}
\normalsize

$M_{fit}$, the mass returned by the fit, is expressed in MeV. The above formula defines a scale-factor-dependent probability that D0 obtained a signal at least as discrepant with $M_{CDF}$ as one observed, and with a normalization of 18.4 events or larger, assuming that the data actually contained a $\Omega_b$ signal with mass equal to that measured by CDF within uncertainties, and rate compatible with the CDF rate. All this can be studied as a function of $k$, the scale factor affecting the 13-MeV systematic uncertainty assigned by D0 to their mass measurement.

\begin{center}
\begin{figure}
\centerline{\epsfig{file=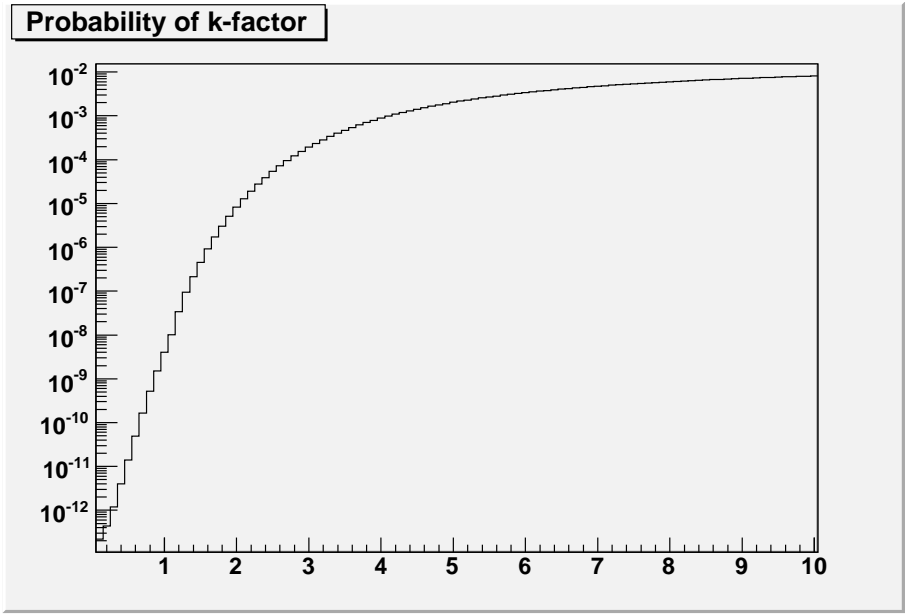,  width=12cm, angle=0, clip=}}
\caption{\em Probability to observe a D0-like signal in data containing a CDF-like one, as a function of the scale factor $k$ applied on the systematic uncertainty affecting the D0 mass measurement.}
\label{f:kfactor}
\end{figure}
\end{center}

The result of the exercise is shown in Fig.~\ref{f:kfactor}. The graph stresses the fact that for $k=1$ the probability that D0 find 18.4 or more signal events at $M=6165 MeV$, given a $\Omega_b$ with mass of $6054.4 \pm 6.9 MeV$ producing $6.0^{+3.95}_{-3.74}$ events in their sample, is very small~\footnote{ Much smaller, in fact, than the one corresponding to the hypothesis that there is no signal anywhere in the spectrum, which we have estimated at $P_{null}=4.1\times 10^{-6}$ in section~\ref{s:d0significance}, because we are now testing a quite different null hypothesis, namely that the two experiments are compatible both in mass and rate. However, it is to be noted that the smoothness of the curve in Fig.~\ref{f:kfactor} is deceiving, since it hides the fact that it is based on just a million-pseudoexperiment test: probabilities below $10^{-6}$ are untrustworthy. }; by blowing up the D0 mass systematics by a factor three, however, the probability rises to about a half-thousandth; a factor $k \sim 6$ is necessary to bring it to three-sigma values. The two mass measurements are not altogether too incompatible, {\em if} we assume that the CDF result is correct {\em and if} we are willing to admit that the D0 systematics were seriously underestimated.

The same exercise just discussed could of course be performed by taking the D0 stand: that is, one might now assume that D0 obtained mass and production rate of $\Omega_b$ baryons right, and verify how likely it is that CDF could get a mass result discrepant with the true value by fitting their mass distribution, as a function of a scale factor multiplying the CDF mass systematics. To circumvent the unavailable information on the mass versus lifetime of all CDF candidates it is in principle possible to rely on the distribution at the bottom of Fig.~\ref{f:omegab_signals}, which is a low-background, 35-event selection used as the basis of the preliminary mass-only fit described in the CDF publication, where they find a smaller-significance signal  corresponding to a $4.9\sigma$ effect according to CDF. This time, however, there is an evident problem of self-consistency of the hypothesis. The signal rate estimated by D0 is almost three times as large as the one estimated by CDF, so one would have to create the pseudo-CDF histograms by inserting $35.6$ signal events, as computed in section~\ref{s:twohyp} above, in a histogram containing a total of 35. The problem is that one cannot ignore that the majority of the 35 events contained in the original CDF histogram are incompatible with being $\Omega_b$ events, regardless of whether $M_{D0}$ or $M_{CDF}$ is assumed to be the correct $\Omega_b$ mass. A more meaningful way to question the accuracy of the CDF result for the sake of bringing the two mass measurements in agreement is described in the next section.


\subsection { The second hypothesis: a D0 mass bias and a CDF rate error \label{s:twosf}}
\vskip .3cm

The observation that rate measurements in both experiments are imprecise suggests one to test a different hypothesis to investigate further the discrepancy. One may study the probability that D0 observed a signal at a mass such that $|M_{D0}-M_{CDF}|>110.6 MeV$, simultaneously as a function of the scale factor $k$ on the D0 mass systematics {\em and} of an independent scale factor $\gamma$ on the $\Omega_b$ signal expected in the D0 histogram given the CDF rate measurement. For $\gamma=1$, $6.0^{+3.95}_{-3.74}$ events should be present in the D0 mass histogram, and from the test of the previous section the case $k=1$ can be judged highly unlikely; however, for a larger value of $\gamma$ the ``spill-over'' effect discussed in section~\ref{s:twohyp} might make the D0 observation more likely.

\begin{center}
\begin{figure}
\centerline{\epsfig{file=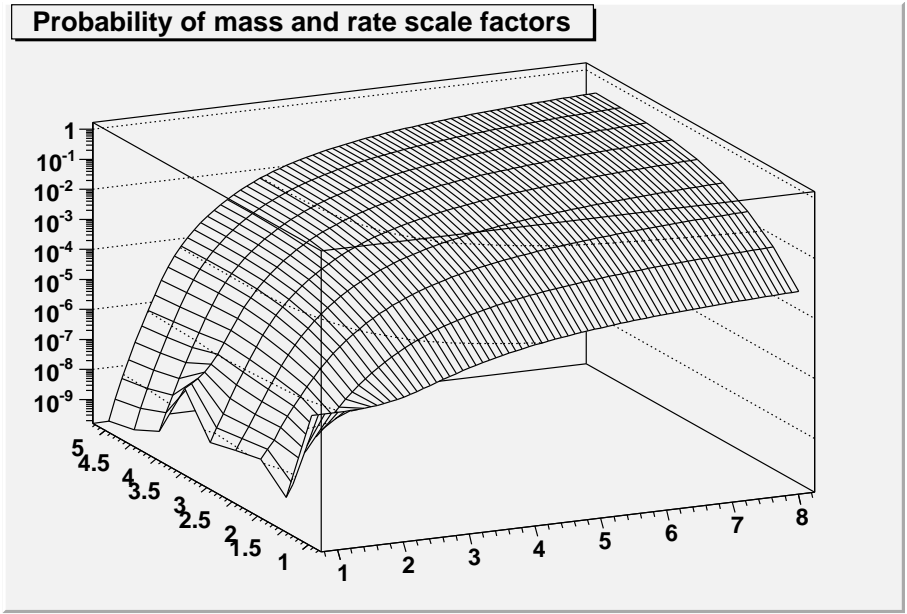,  width=12cm, angle=0, clip=}}
\caption{\em Probability to observe a D0-like signal in data containing a CDF-like one, as a function of the scale factor $k$ applied on the systematic uncertainty affecting the D0 mass measurement (running from 1.0 to 8.0 on the x-axis), and as a function of the scale factor $\gamma$ applied on the rate of $\Omega_b$ candidates expected in the D0 histogram by assuming the CDF measured rate (running from 1.0 to 5.0 on the y axis). See the text for caveats and details.}
\label{f:twoscalefactors}
\end{figure}
\end{center}

The test is performed by choosing nine values from 1.0 to 5.0 of the rate scale factor $\gamma$, and constructing for each of them 500,000 pseudo-histograms in which the mass of  $N_{\Omega_b}= 6.0 \times \gamma$ entries is sampled from the CDF $\Omega_b$ mass measurement ($M_{CDF}=6054.4 \pm 6.9 MeV$) with a $34 MeV$ resolution, and the remaining $79-N_{\Omega_b}$ entries are obtained from a uniform distribution. Along with the nine values of $\gamma$, 100 values of the scale factor $k$ are sampled from 0.1 to 10.0~\footnote {As for the test of section~\ref{s:d0massbias}, there is no need to re-generate pseudo-histograms to test different values of $k$: one just needs to compute 100 different values of the integral for each pseudoexperiment returning 18.4 or more signal events. For $\gamma$, instead, different sets of pseudoexperiments are needed.}; the calculation of probability of the D0 observation follows the method described in the previous section. Results are presented in Fig.~\ref{f:twoscalefactors}; a selection of numerical results is also provided in the table below.

\begin{center}
\begin{tabular}{c|c|c|c}
D0 mass syst. scale factor & CDF rate scale factor & Equivalent CDF rate bias & Estimated compatibility\\
$k$              & $\gamma$    & $N(\sigma) = (\gamma-1)/0.45$ & P \\
\hline
1.0 & 1.0 & 0.  &  $3.2 \times 10^{-6}$ \\
2.0 & 1.0 & 0.  &  $4.4 \times 10^{-6}$ \\
3.0 & 1.0 & 0.  &  $2.1 \times 10^{-5}$ \\
4.0 & 1.0 & 0.  &  $7.5 \times 10^{-5}$ \\
5.0 & 1.0 & 0.  &  $1.6 \times 10^{-4}$ \\
1.0 & 2.0 & 2.2 &  $4.8 \times 10^{-8}$ \\
2.0 & 2.0 & 2.2 &  $3.3 \times 10^{-5}$ \\
3.0 & 2.0 & 2.2 &  $7.6 \times 10^{-4}$ \\
4.0 & 2.0 & 2.2 &  $3.5 \times 10^{-3}$ \\
5.0 & 2.0 & 2.2 &  $8.0 \times 10^{-3}$ \\
1.0 & 3.0 & 4.4 &  $3.9 \times 10^{-8}$ \\
2.0 & 3.0 & 4.4 &  $1.4 \times 10^{-4}$ \\
3.0 & 3.0 & 4.4 &  $4.2 \times 10^{-3}$ \\
4.0 & 3.0 & 4.4 &  $2.0 \times 10^{-2}$ \\
5.0 & 3.0 & 4.4 &  $4.8 \times 10^{-2}$ \\
\hline
\end{tabular}
\end{center}

Before we try to interpret the above numbers, a few points must be made about their determination.

\begin{itemize}

\item The two scale factors $k$ and $\gamma$ have a quite different nature: the first refers to an increase of the systematic error in the D0 mass measurement, while the second is a factor scaling up the effective rate of $\Omega_b$ baryons with respect to the CDF measurement. Since the latter is measured with a 45\% relative uncertainty, a rate scale factor $\gamma=3.0$ equates to a CDF underestimate of the rate by 4.4 standard deviations, as indicated in the third column. 

\item Another thing to note about these numbers is that they do not directly compare to the probability extracted with background-only pseudoexperiments discussed in section~\ref{s:d0significance} ($P_{null}=4.1 \pm 0.8 \times 10^{-6}$). The reason is that we have been testing two very different questions. In the test of the null hypothesis of section~\ref{s:d0significance} one searched anywhere in the mass spectrum for a signal  with a normalization exceeding 18.4 events, among 79 mass values distributed uniformly; here, instead, we refer to a situation where a signal is present, and try to determine how likely it is that a signal is fit at a mass at least $110.6 MeV$ away from the one measured by CDF. The numbers indicate that the D0 and CDF mass values are so distant that a leak of signal events from the ``true'' mass (assumed to be $M_{CDF}$) to the vicinities of $M_{D0}$ is unable, for $\gamma=1$, to make up for the concurrent effect: the Gaussian degrees of freedom are used, in the vast majority of cases, to accommodate the few mass values really generated at $M_{CDF}$, depleting the chance of a fluctuation where D0 sees its signal.

\item The method used to compute probabilities, based on the integration of Gaussian functions, provides a certain level of smoothness in our estimates of very small probabilities, but it does not protect them from the large random fluctuations intrinsic of the pseudoexperiments. This is evident by observing the erratic behavior of the part of the surface in~\ref{f:twoscalefactors} corresponding to small $k$: a few pseudoexperiments fluctuating to yield a signal above 18.4 events at a mass close to $M_{D0}$ among the 500,000 generated for $\gamma=1.0$, and similarly for the set generated for $\gamma=3.5$, produce steps in a surface whose real shape should be smooth and monotonous.

\end{itemize}

The numbers presented in the table clearly show that a decrease of the $6.2 \sigma$ discrepancy between the CDF and D0 mass measurements to a less-than-$3\sigma$ effect can be obtained by several combinations of causes. We list two of them below, which join the simpler $k>1, \gamma=1$ cases to which the tests described in section~\ref{s:d0massbias} correspond.

\begin{enumerate} 
\item A choice of $k=3$ and $\gamma=3$ brings the observed discrepancy to an estimated probability of $4.2 \times 10^{-3}$;
\item If one only accepts that the CDF rate measurement is underestimated by a factor of 2 (a $2.2\sigma$ effect, if one believes the CDF rate uncertainty), then a D0 mass systematics SF $k=4.0$ is again sufficient to bring the probability of the observed discrepancy to $3.5 \times 10^{-3}$.
\end{enumerate}

Of course, a numerical analysis such as the one above is useless if it is not complemented by physical insight based on hard facts: how the analyses were performed, what could have gone wrong, what implicit biases affect the experimental situation. It is the opinion of the author that the most likely hypothesis for the observed conflict of mass measurements of the $\Omega_b$ baryon is indeed the conspiracy of a combination of factors: those examined above, but also others, which are easily overlooked because they are of exogenous nature. One of them might be the fact that CDF performed its measurement {\em after} D0 found a signal at a mass discrepant with the most credited predictions for the sought baryon. It is likely that this fact led CDF to choose the most robust method to measure the $\Omega_b$ mass: their selection of the data is based on straight cuts rather than advanced analysis techniques, for the declared purpose of retaining the possibility to calibrate the mass reconstruction of the $\Omega_b$ to that of well-known lighter baryons and mesons yielding a similar decay topology, which are retained by the conservative selection. CDF thus apparently traded in a less-than-optimal rate measurement for the best possible mass measurement. If that is the case, a discrepancy in the mass between D0 and CDF --fostered by a sub-optimal rate uncertainty in the CDF measurement-- becomes more likely.

\section{Conclusions}
\vskip .5cm

The recent mass estimates of the $\Omega_b$ baryon obtained by the D0 and CDF collaborations in their Run II datasets disagree by more than six standard deviations. The extracted signals which are the basis of those estimates are also quite unlikely to be due to statistical fluctuations of backgrounds. The inconsistency calls for more investigations, which the Tevatron experiments will no doubt produce in the near future. In the meantime, an assessment of the situation from a statistical standpoint is called for.

In this paper we offer the results of a study of the compatibility of the D0 and CDF results, performed to try and quantify their apparent inconsistency with several simple tests. A check of the calculation reported in the D0 publication demonstrated that the significance quoted by the D0 collaboration is overestimated, both because of an inconsistent calculation and because of their neglecting a trials factor. The possibility of underestimated systematic uncertainties in the D0 mass measurements is then considered with pseudoexperiments, in combination with a hypothesis that the CDF measured yield of $\Omega_b$ decays is underestimated. By assuming the CDF mass measurement is correct, it is proven that the likelihood of a D0 result as discrepant and as significant as the one seen also depends on the rate bias assumed in the CDF result. As a result of the calculations presented in this paper, a viable hypothesis can be put forth: if mass systematics quoted by D0 are inflated by a factor of three and the signal rate measured by CDF is inflated by a similar amount, the apparent 6.2-standard-deviation discrepancy between the CDF and D0 results gets reduced to a less-than $3 \sigma$ effect. 

At the time of writing, the Tevatron collider is just about to cross the mark of 7 inverse femtobarns of proton-antiproton collisions delivered to the two experiments; a five-fold increase in statistics for the D0 analysis is thus possible. It is therefore only a matter of time before the apparent inconsistency between the two observations of the $\Omega_b$ baryon is resolved.



\end{document}